\documentclass[a4paper,fleqn]{cas-dc}

\usepackage[numbers]{natbib}

\usepackage{CJKutf8}
\usepackage{bbm}
\usepackage{caption}
\usepackage{subcaption}
\usepackage{graphicx}
\usepackage[capitalise]{cleveref}
\usepackage{soul}
\usepackage{hyperref}

\def\tsc#1{\csdef{#1}{\textsc{\lowercase{#1}}\xspace}}
\tsc{WGM}
\tsc{QE}
\tsc{EP}
\tsc{PMS}
\tsc{BEC}
\tsc{DE}

\begin{document}
\begin{CJK}{UTF8}{gbsn}
\let\WriteBookmarks\relax
\def\floatpagepagefraction{1}
\def\textpagefraction{.001}
\renewcommand{\figurename}{Fig.}

\shorttitle{EAR}

\shortauthors{L. Tan et~al.}

\title [mode = title]{EAR: Edge-Aware Reconstruction of 3-D vertebrae structures from bi-planar X-ray images}                      



%
\author[1]{Lixing Tan}[style=Chinese,orcid=0000-0002-0323-8802]
\credit{Conceptualization, Methodology, Software, Validation, Formal
analysis, Investigation, Writing \& original draft, Visualization}
\author[1]{Shuang Song}[style=chinese]
\credit{Conceptualization, Methodology, Formal
analysis, Investigation, Writing \& review \& editing, Supervision, Visualization}
\author[2]{Yaofeng He}[style=chinese]
\credit{Validation, Formal Analysis}
\author[3]{Kangneng Zhou}[style=chinese]
\credit{Formal Analysis}
\author[4]{Tong Lu}[style=chinese]
\credit{Resources}
\author[1,2]{Ruoxiu Xiao}[style=Chinese]
\cormark[1]
\credit{Conceptualization, Resources, Project administration, Funding acquisition}

\affiliation[1]{organization={School of Computer and Communication Engineering},
    addressline={University of Science and Technology Beijing}, 
    postcode={100083 Beijing}, 
    country={China}}
\affiliation[3]{organization={College of Computer Science}, addressline={Nankai University},
    postcode={300000 Tianjin}, 
    country={China}}
\affiliation[2]{organization={Shunde Innovation School}, addressline={University of Science and Technology Beijing},
    postcode={100024 Foshan}, 
    country={China}}
\affiliation[4]{organization={Visual 3-D Medical Science and Technology Development Co. Ltd,},
    postcode={100070 Beijing}, 
    country={China}}

\cortext[1]{Lixing Tan and Shuang Song contributes equally.}
\cortext[cor1]{Corresponding author at: The School of Computer and Communication Engineering, University of Science and Technology Beijing, Beijing 100083, China. E-mail address: \href{xiaoruoxiu@ustb.edu.cn}{xiaoruoxiu@ustb.edu.cn}}

\begin{abstract}
X-ray images ease the diagnosis and treatment process due to their rapid imaging speed and high resolution. However, due to the projection process of X-ray imaging, much spatial information has been lost. To accurately provide efficient spinal morphological and structural information, reconstructing the 3-D structures of the spine from the 2-D X-ray images is essential. It is challenging for current reconstruction methods to preserve the edge information and local shapes of the asymmetrical vertebrae structures. In this study, we propose a new Edge-Aware Reconstruction network (EAR) to focus on the performance improvement of the edge information and vertebrae shapes. In our network, by using the auto-encoder architecture as the backbone, the edge attention module and frequency enhancement module are proposed to strengthen the perception of the edge reconstruction. Meanwhile, we also combine four loss terms, including reconstruction loss, edge loss, frequency loss and projection loss. The proposed method is evaluated using three publicly accessible datasets and compared with four state-of-the-art models. The proposed method is superior to other methods and achieves $25.32\%$, $15.32\%$, $86.44\%$, $80.13\%$, $23.7612$ and $0.3014$ with regard to MSE, MAE, Dice, SSIM, PSNR and frequency distance. Due to the end-to-end and accurate reconstruction process, EAR can provide sufficient 3-D spatial information and precise preoperative surgical planning guidance.

\end{abstract}




\maketitle

\section{Introduction}

As the main structure supporting the human body, spine is of great significance in providing stability and flexibility to body posture and enabling coordinated movement. Spine-related diseases, including scoliosis, spinal dysraphism, and intervertebral disc protrusion, are greatly threatening human health. The misdiagnosis or delayed treatment can lead to physical deformities and even cardiovascular and pulmonary abnormalities, and further significantly impact the quality and daily functioning of patients. Compared with natural images, medical images contain higher ambiguity and complexity \cite{liu2016medical}. As the imaging gold standard of spinal disease diagnosis and treatment, X-ray images can visualize bone structures, locations, and abnormalities with fast imaging speed, high resolution and low radiation dosage. However, due to the projection from 3-D space to 2-D space is a dimension reduction process, much spatial information has been lost \cite{zhao2018efficient}, which makes it difficult for the surgeons to accurately measure the spinal morphology parameters, such as the spinal curvature, rotation and local deformation. In addition, the asymmetric structural distribution and occlusion also increase the difficulty of the reconstruction. Though computed tomography (CT) can clearly present the detailed spinal structures from the 3-D views, the imaging speed is slow. Meanwhile, due to the high doses of radiation, unavoidable harm will be caused to both the patients and clinicians. Hence, it is highly necessary to reconstruct the 3-D spine from the X-ray images with limited viewing angles. 

Given that much information is lost during the projecting procedure of X-ray images, it is an ill-posed problem to reconstruct the 3-D volume structure from 2-D projections. The traditional solutions mainly include the kriging-based \cite{mitton2000, humbert2009} and statistical shape model (SSM)-based \cite{GuoyanZheng2011, boisvert2008} methods. These methods are motivated by the deformation of 3-D flexible structures observed in consistent 2-D projections of X-ray images in specific imaging views. However, these methods often require additional prior information, such as anatomical structures.

Recently, along with the explosive development of deep learning models, 3-D reconstruction can be realized by directly learning to map the distribution of the 2-D X-rays to the distribution of the 3-D volumes. While the possibility and advantages of reconstruction with deep learning-based approaches have been demonstrated, drawbacks still exist and can be summarized in three aspects: (1) due to the dimension difference between the input and output images, there is a considerable semantic gap between the feature maps; (2) Reconstruction relies on feature matching, necessitating a balance between accuracy and real-time performance; (3) To improve the accuracy of the reconstruction, related researchers focus on disentangling spatial transformation and content information. It generally results in lower accuracy within edge regions. 

In this study, to relieve the above-mentioned problems, we proposed the Edge-Aware Reconstruction network (EAR) which is based on the auto-encoder architecture. During data preparation, we first expanded the dimension by replicating the bi-planar X-ray images along both anterior-posterior (AP) and lateral (Lat) directions. In such a way, the reconstruction can be realized with 3-D convolution by learning the potential correlations of features in the higher dimension. In the proposed network, we use the symmetrical auto-encoder architecture with five layers as our backbone. Convolutional block attention module (CBAM) is embedded in the shortcuts between the symmetrical downsampling and upsampling layers to increase the weights utilizing the spatial and channel attention mechanisms. In addition, to ameliorate the precision of edge reconstruction, we took advantage of feature similarity in the frequency domain and designed the edge attention module (EAM) and frequency enhancement module (FEM) to further strengthen the weights of edge structures. In our loss function, we used the edge displacement as an auxiliary constraint to enlarge the influence of the edge region. Moreover, we utilized the maximum intensity projections (MIP) of the outcome as the additional shape to speed up the training and improve the edge shape consistency. Finally, we defined a frequency distance (FD) loss term to convert the reconstruction results into the frequency domain and thus gain improved quality of high-frequency components corresponding to the edge and sharp regions.

The contributions of this study are threefold as follows:

\begin{itemize}
    \item We propose a novel and effective edge-aware reconstruction network by combining the frequency enhancement module and edge attention module, improving the accuracy of 3-D reconstruction from X-ray. The proposed EAR can not only handle the detailed parts of the vertebral body and spinous process, but also provides an end-to-end way to reconstruct the asymmetrical vertebrae by preserving the edge information and local shapes.

    \item The frequency enhancement module transforms the feature maps into the frequency domain. It aggregates the high-frequency information corresponding to edge structures and eases the perception ability of the network to the edge regions. Additionally, the design of frequency distance loss further enhances the reconstruction accuracy of high-frequency details.

    \item We provide a general way to generate the X-ray images from different view angles, which can provide more databases to train and evaluate the reconstruction networks.

    \item We rigorously evaluate the proposed method both quantitatively and qualitatively using three public spine datasets, with the comparison of four other state-of-the-art 3-D reconstruction methods. The results show that our proposed EAR can not only effectively reconstruct the vertebrae body but also preserve the detailed edge structures.


\end{itemize}

\section{Related Work}
The traditional methods for 3-D reconstruction from X-ray radiographic images are based on the deformation models. Generally, a 3-D deformable reference structure is firstly pre-defined. By transforming the reference structure with an affine matrix, the 3-D reconstruction can then be realized when the errors between the projected images of the transformed reference structure and the target 2-D images reach the minimum value. The earlier methods are mainly derived from the kriging- and SSM-based approaches \cite{GuoyanZheng2011,boisvert2008,benameur2003,benameur2005}, which assume that the spine is a rigid structure and always with similar morphologies. In the kriging-based method, a series of anatomical landmarks need to be selected in the X-ray images. By calibrating the corresponding landmarks from different X-ray images, the 3-D locations of these landmarks are obtained. Then the landmarks in the 3-D space are sent into an interpolation process to generate detailed points of the target model. Such kinds of methods can be optimized by combining with multi-view geometry techniques. Mitton et al. performed deformation on an elastic object so that the stereo-corresponding and non-stereo-corresponding observations in different views can be maintained \cite{mitton2000}. This technique requires the annotation of representative anatomical landmarks on the projections, and the reconstructed edge regions exhibit significant distortion. Furthermore, this method demonstrated a significantly slow reconstruction speed, which cannot meet the real-time requirements of surgical procedures. Humbert et al. constructed a parametric model based on longitudinal and transversal inferences for fast reconstruction \cite{humbert2009}. Generally, to personalize the reconstruction function, large amounts of parameters need to be defined and optimized which greatly decreases the model efficiency. To reduce the influence of the difference between the pre-defined deformable structure and the target reconstructed object to reconstruction accuracy, a mean statistical shape of spines is spatially deformed to realize the reconstruction. The reconstruction process is to find a series of optimal parameters by maximizing the similarity between the projected structures of the transformed images and real 2-D spines. Benameur et al. \cite{benameur2003} modeled the prior geometric structure of each vertebra and realized the 3-D reconstruction by using the first mode of variation in the Karhunen-Loeve expansion. In their later study \cite{benameur2005}, they divided the reconstruction into two stages which include a global deformation of the whole spine and a local deformation of each vertebra. The global one is based on statistical modal analysis while the local one is described by the first-order Markov. A stochastic optimization algorithm is then utilized to solve the two independent minimization processes. Boisvert et al. \cite{boisvert2008} included the 3-D positions of landmarks in the local coordinate system to capture the scoliosis of spines. However, due to the sparse estimation, global distortions tend to occur with regard to the vertebrae morphology. Considering the intuitive appearance reconstruction of spines, Zheng et al. \cite{GuoyanZheng2011} proposed the point distribution model to build the SSM. By computing the deformable field between the point cloud from input 2-D images and the SSM using an iterative closet point algorithm, the correspondences between 2-D and 3-D spaces can be obtained. The thin-plate splines method is then utilized to deform the shape of SSM.



Due to traditional reconstruction methods depend on the camera parameters of capturing X-ray images, it is much more complicated and time inefficient to iteratively solve the deformation attributes. Additionally, the deformation-based methods are more suitable for the mesh modeling instead of the volume images.

\begin{figure*}[tbp]
    \centering
    \includegraphics[width=\textwidth]{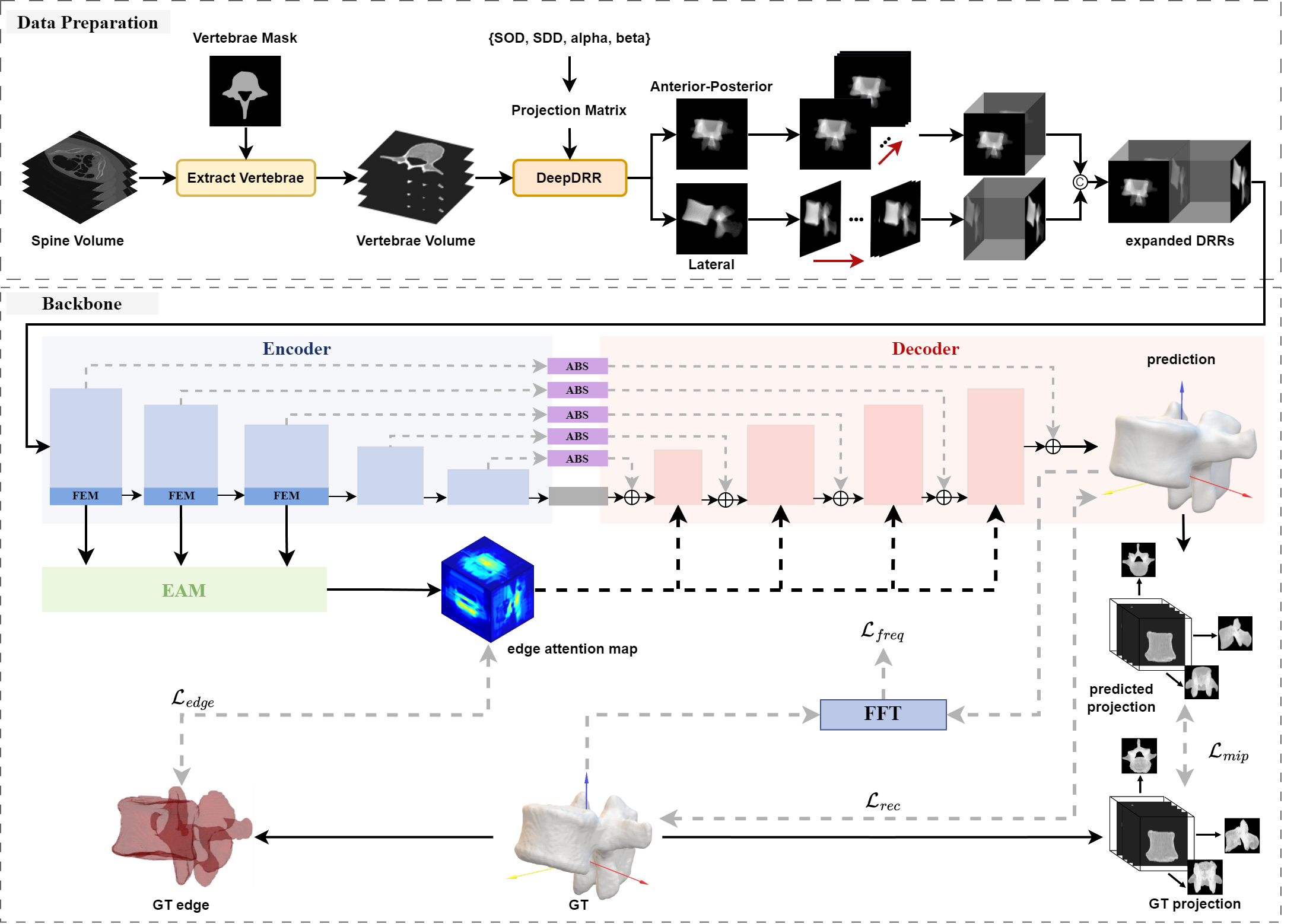}
    \caption{Overview of the proposed EAR for vertebrae reconstruction: an encoder, a frequency enhancement module, an edge attention module, an attention-based shortcut and a decoder.}
    \label{figure1}
\end{figure*}

To ease above mentioned questions, solutions to 3-D reconstruction increasingly employ deep learning methods \cite{PhilippHenzler2018,LiyueShen2019,XingdeYing2019,RongjunGe2022,YoniKasten2020,MdAminurRabRatul2021,LingJiang2021,ZheyeChen2023,zhou2023mate3d}. Kim et al. \cite{HangkeeKim2019} extracted features from convolutional neural network (CNN) and then combined with the SSM to optimize the reconstruction process. Henzler et al. \cite{PhilippHenzler2018} first attempted to reconstruct the 3-D volumes under the auto-encoder architecture. Since the proposed method is based on the single-view image, it is difficult to restore comprehensive structures and also tends to introduce ambiguities. Shen et al. proposed a three-layer Patient-Specific Reconstruction (PSR) to reconstruct a CT volume from X-ray images \cite{LiyueShen2019}. In this framework, they demonstrated the possibility of reconstructing CT images from single-planar X-ray images, and their experiments demonstrated that the results could be further refined by increasing the number of views. However, PSR has some limitations: due to the deep network architecture and its pure CNN backbone instead of auto-encoder backbone, the training of the network requires a substantial amount of GPU memory and training time. Furthermore, due to the absence of additional prior information, the network can only differentiate between soft tissues and bones based on intensity, which increases the complexity of the reconstruction. In comparison, X2CT-GAN designed a generative adversarial network to fuse information from two orthogonal X-rays so that the dimension of the data can increase from 2-D to 3-D \cite{XingdeYing2019}. Katen et al. proposed the dimensional expansion scheme by replicating the bi-planar X-ray images along the orthogonal axes and further fed into the auto-encoder network. Since the operation is in the 3-D space, the pre-processing step greatly reduces the reconstruction difficulty of feature mismatching within cross-dimension.

Deep learning-based approaches have potentially made it possible to reconstruct the whole structures with complicated morphology. In order to pay more attention to local shape reconstruction, attention mechanism is gradually becoming an essential part of the networks. The principle of attention mechanisms is to emulate the human visual cognition system by assigning distinct weights to target regions. Such strategy is capable of catching complex semantic relationships in medical images of multi-modalities \cite{zhang2019net, Fan2020, xie2021, Wang2022a, Xia2022, chen2023}. To fuse the features from multi-scale stage, Chen et al. \cite{ZheyeChen2023} introduced a feature attention guidance module (FFAG) to aggregate image features from multi-resolutions. Considering the asymmetric shape and edge sharpening of the whole spine, Zhang et al. \cite{zhang2019net} implemented an edge attention module to guide the segmentation process and gained significant accuracy. Fan et al. utilized reverse attention to combine current feature maps and edge attention maps \cite{Fan2020}. By regarding the feature map from low-level convolution as edge attention, 
they concatenated it with multi-resolution feature maps through multi-scale down-sampling, thus implicitly learning edge features. Wang et al. \cite{Wang2022a} proposed the edge attention preservation module to improve the perceptibility of the network to the edge regions. Xia et al. \cite{Xia2022} introduced the reverse edge attention module and an edge-reinforced loss term to jointly enhance the voxel weights close to the edges. Though spatial edge information can be preserved, research concerning edge reconstruction is still scarce.

\section{Methodology}
Fig. 1 shows the overall workflow of the proposed 3-D reconstruction framework. In our framework, the dataset after dimension expansion is extended to two volumes with the shape of $128\times 128\times 128$ so that the dimensional gap between 2-D X-rays and 3-D volumes is minimized. The proposed pipeline increases the gap between high- and low-frequency signals and employs an edge-awareness decoding process. The ABS improves the reconstruction quality by suppressing the discrete noise region. 

\begin{figure*}[htbp]
    \centering
    \includegraphics[width=\textwidth]{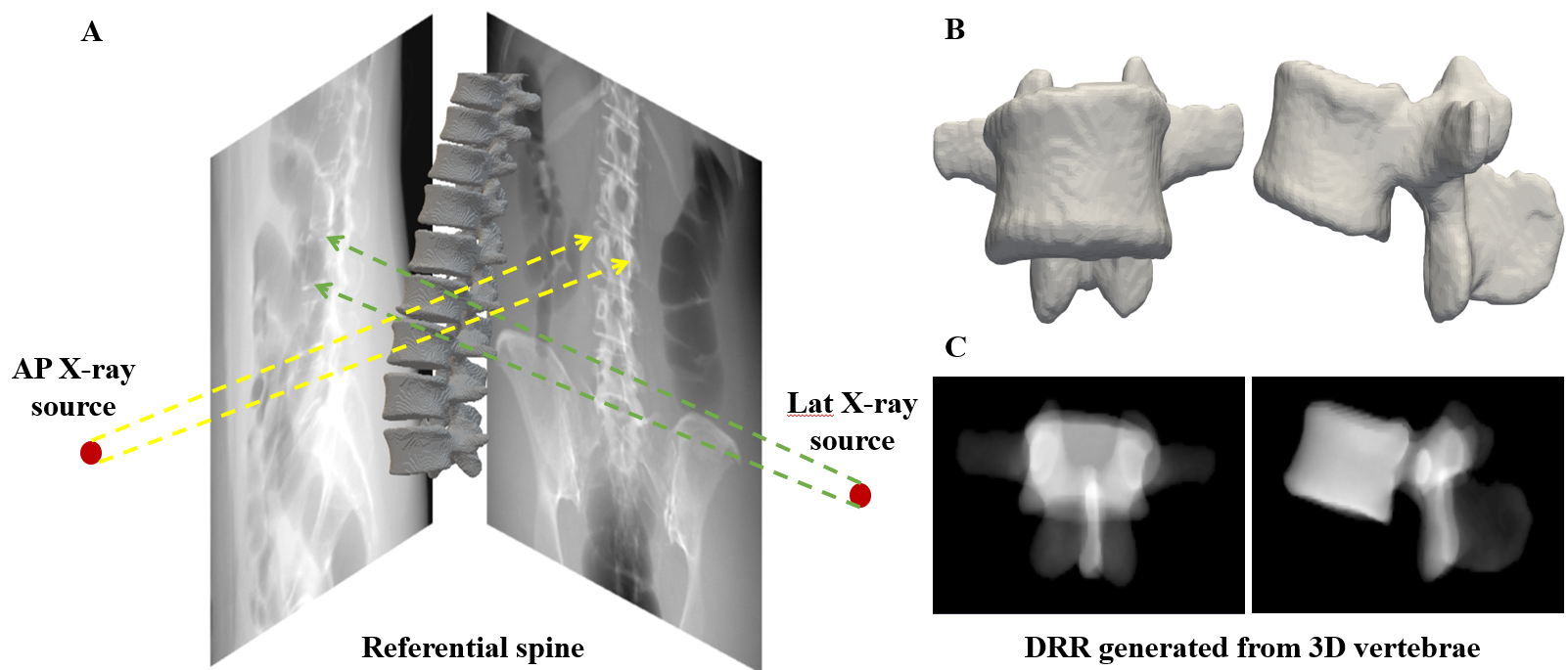}
    \caption{Bi-planar X-rays for 3-D reconstruction (A). The referential spine is displayed in 3-D view and in both anterior-posterior (AP) and lateral (Lat) projections. Single vertebrae models (B). Digitally reconstructed radiographs generated from 3-D models (C) showing the simulated AP and Lat views of the vertebrae.}
    \label{fig_drr}
\end{figure*}

\subsection{Data Preparation}

Considering the limitations of imaging system, it is difficult to capture the X-ray spine images from multi-views in a single imaging procedure. Compared to the work from Ying et al. who synthesized the X-ray images from CT scans while using CycleGAN to map the real X-rays to the synthesized style to minimize the gap between synthesized images and real X-rays \cite{XingdeYing2019}, we directly generated much more realistic X-ray images of multi-views by modifying DeepDRR \cite{gao2022} by which paired CT and X-rays of the same patient can be made to facilitate the reconstructions of CT scans from multi-view X-ray images using the deep networks. The illustration of digitally reconstructed radiograph (DRR) is shown in \cref{fig_drr}. In the generation, we resampled the CT to a resolution of 1 mm $\times $ 1 mm $\times $ 1 mm and then extracted single vertebrae from the uniformly resampled CTs using masks and constructed the projection matrix with variable Euler angles (alpha and beta) and fixed distances, both source to object distance (SOD) and source to detector distance (SDD). As for the attributes of DeepDRR, we set the energy spectrum to '60KV\_35AL' and used more than 1000000 photons to simulate the realistic scatter process. We focus on the single vertebra reconstruction instead of the whole spine (12 thoracic vertebrae and 5 lumbar vertebrae) considering the limited GPU memory. The whole spine can then be composed by reconstructing each vertebrae respectively. 

All volumetric scans are resized to $128\times128\times128$ and the 2-D projections are resized to a resolution of $128\times 128$ due to the limitation of memory and the purpose of computational efficiency. To improve convergence, a z-score normalization step was employed where the intensities of the volumetric data were manipulated to have a zero mean and a unit variance.

\subsection{Backbone Network of 3-D Reconstruction}

The backbone network utilizes the typical U-Net structure with five layers. In the encoder part of the network, the images with enhanced dimensions are regarded as the input. In each layer of the down-sampling path, the module begins with a convolution operation with a kernel size of $3\times 3\times 3$ and a stride of 2. Afterward, it adds a convolution operation with a kernel size of $3\times 3\times 3$ and a stride of 1. Then a batch normalization (BN) and a leaky rectified linear unit (leaky ReLU) follow the convolution operation. At the end of the first three layers in the encoder, we added a simple yet effective frequency enhancement module (FEM) to better discriminate the high- and low-frequency contents of the feature maps.

In comparison to the encoder, the decoder performs an up-sampling operation on the feature maps and restores the resolution of the feature maps layer by layer. 
In the decoder, the designed layer consists of a transposed convolution layer with a kernel size of $2\times 2\times 2$ and a stride of 2 and a convolution layer with the same setting as the second convolution operation used in the down-sampling layers. In addition, due to the lack of edge information, we introduced 3-D edge attention module (EAM) into each decoding layer to retain more edge information. Lastly, we designed an attention-based shortcut (ABS) to get a more fine-grained outcome.

In the next sections, we will look at the two aspects: edge enhancement (in sections 3.3.1 and 3.3.2) and optimized shortcut connection (in section 3.4).


\subsection{Edge-aware Attention Module}

Considering the irregular and asymmetrical shape of the vertebrae, the most difficult part of reconstructing the vertebrae is to accurately preserve the edge information. Hence, in the reconstruction process, to better perceive the edge information, we designed the FEM and further combined it with the 3-D EAM to improve the attention to the edge regions. 


\subsubsection{Frequency Enhancement Module}
\begin{figure}[htbp]
    \centering
    \includegraphics[width=0.5\linewidth]{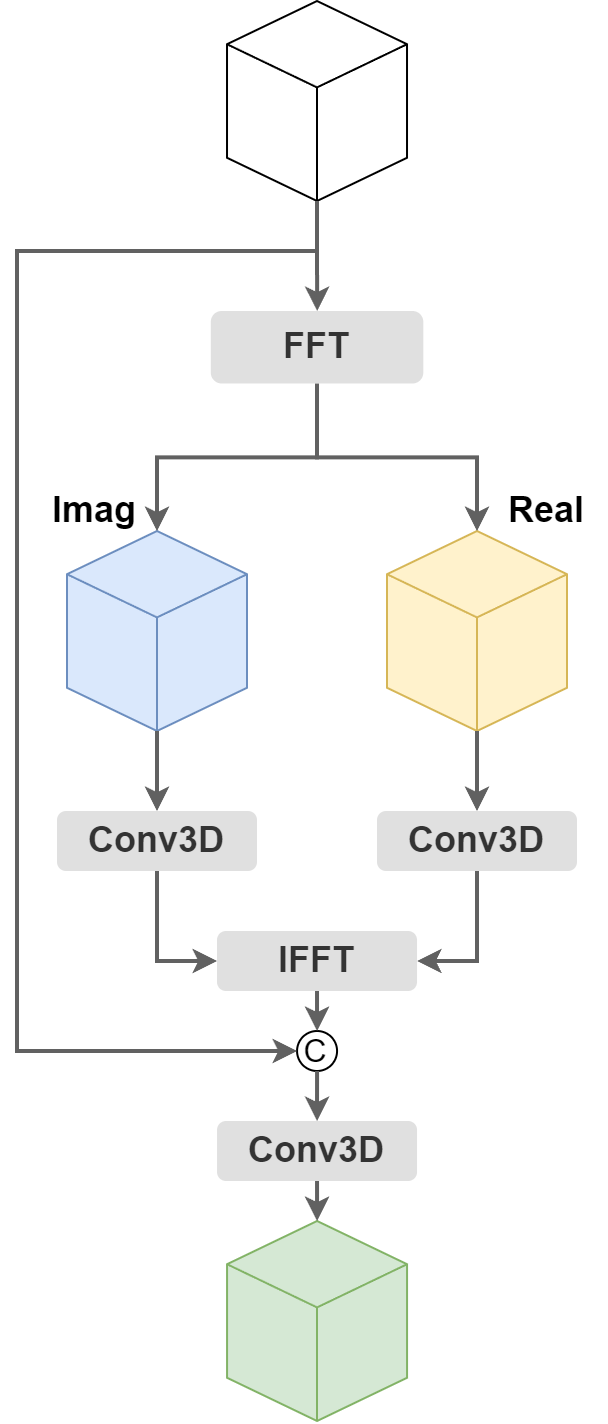}
    \caption{The diagram of the frequency enhancement module. FFT denotes the fast Fourier transform. Imaginary and real parts after performing FFT are denoted as Imag and Real, respectively. IFFT denotes the inverse fast Fourier transform.  }
    \label{figure4}
\end{figure}

Considering that the receptive field in the image domain is limited, the features of edge regions cannot be aggregated and differentiated from other flattened regions. On the contrary, the edge regions in the frequency domain all gather in the high-frequency part which makes it much easier to pay more attention to the edge regions. Hence, by applying the Fourier transform to the feature maps in the encoder path, we design the FEM to emphasize the edge regions, as shown in \cref{figure4}.

In the architecture of this module, the feature maps are first fed into the instance normalization (IN) layer to focus on the statistical property instead of over-smoothing the specificity of the feature maps. Next, a fast Fourier transform (FFT) layer is connected to transform the normalized feature maps into the frequency domain. In this procedure, the feature maps are split into the real and imaginary parts in the frequency domain. Subsequently, two parallel convolution layers are conducted to the real and imaginary parts, respectively. The magnitude of their output complex values from the two layers are then computed by an inverse fast Fourier transform (IFFT). To better complement the features both from the image and frequency domains, the input feature maps to this module are concatenated with the one processed by the frequency layers and further fed into a convolution block (Conv3-D+IN+ReLU).

\begin{equation}
\begin{aligned}
    \mathcal R,\mathcal I &=Conv(FFT(IN(\mathbf{F}_l)))\\
    \mathbf{F}_{FEM} &= ReLU(IN(Conv(IFFT(\mathcal R + i\mathcal I)\oplus \mathbf F_l)))
\end{aligned}
\end{equation}

\subsubsection{3-D Edge Attention Module}

In the process of feature transferring to the decoder part, the edge information will be gradually weakened due to the down-sampling. To force the edge information to provide more constraints to the reconstruction procedure \cite{Fan2020,Wang2022a,Xia2022,ge2023unsupervised}, we designed a 3-D EAM to capture the edge information from both image and frequency domains. 

\begin{figure*}[htbp]
    \centering
    \includegraphics[width=0.9\textwidth]{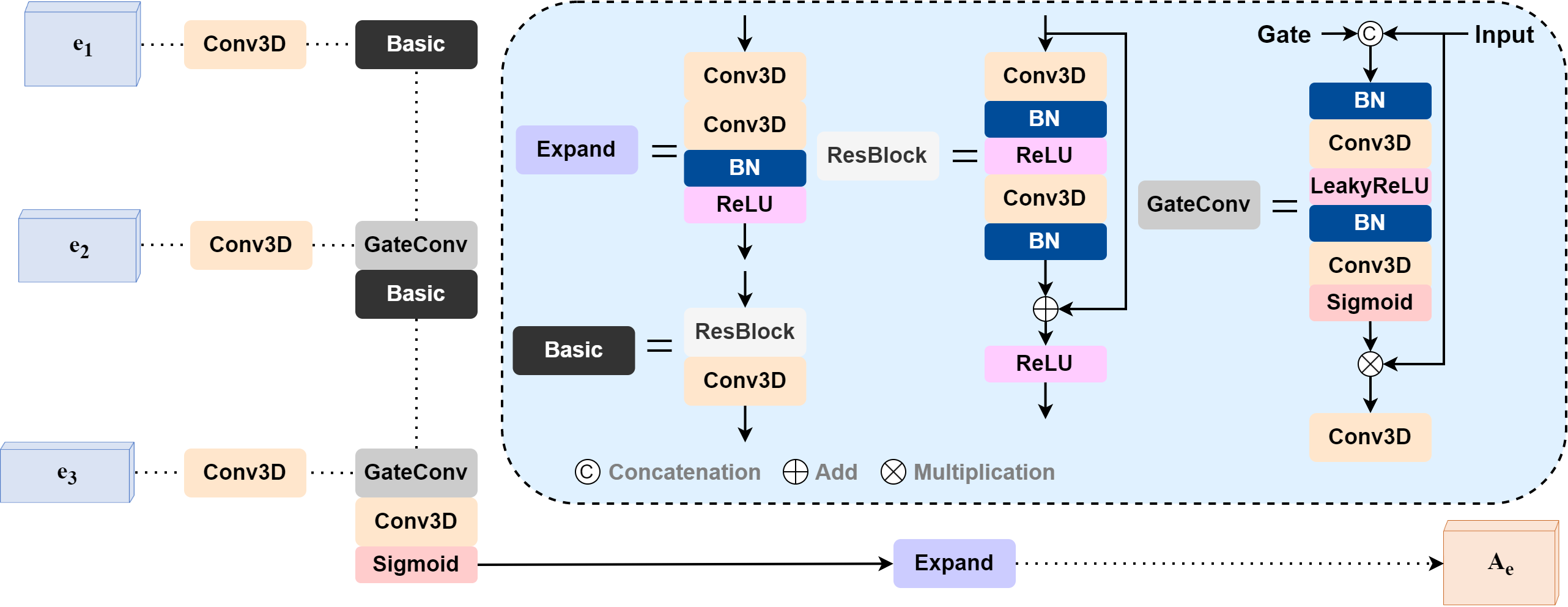}
    \caption{The diagram of the edge attention module (EAM). The features from different layers of the encoder, denoted as $e_1$, $e_2$ and $e_3$, respectively are input to EAM to generate an output attention map $A_e$.}
    \label{fig_eam}
\end{figure*}

In the 3-D EAM, the low-level features extracted by the frequency enhancement module are regarded as the inputs. Inspired by the EANet \cite{Wang2022a}, this module is constituted of the residual blocks comprising the modified gated convolution blocks (m-GCB), as shown in \cref{fig_eam}. In the residual blocks, two convolution blocks (Conv3-D+BN+ReLU) are connected. Gated convolution (GC) was first proposed for the image inpainting task \cite{Yu_2019_ICCV} which distinguishes between the internal and external content of a specified section and  was proved efficient in medical tasks \cite{chen2023ldanet}. By emphasizing the external content over internal redundancy, the weight of edge regions is increased. The m-GCB aims to constrain the network to pay more attention to edge information. In this way, feature maps from the current level and the previous level are concatenated and then fed into two connected convolution blocks to compute the gated maps. In the second convolution block, we use the Sigmoid function as the activation function to capture more edge information. After obtaining the gated maps, the features from the current level are multiplied by the gated maps, enabling the edge attention module to focus solely on the edge information.

Given feature map $\mathbf{F}\in \mathbbm R^{C\times H\times W\times D}$ and its gating vector $\mathbf{F}_g\in \mathbbm R^{C'\times H'\times W'\times D'}$, the GC can be calculated as follow:
\begin{equation}
    \alpha = ((\mathbf{F}\times \sigma(BN(\mathbf{F}\oplus\mathbf{F}_g)))+\mathbf{F})
\end{equation}
where $\oplus$ denotes concatenation, $\sigma$ refers to sigmoid function.

\subsection{Attention-based Shortcut}

Although dimensional enhancement greatly reduces the dimension gap between 2-D X-ray images and 3-D CT volumes, we noticed discrete parts appearing in the outcome due to redundant information generated during dimensional enhancement. To suppress the effect of redundant noise and focus on the bone regions' features, we introduced CBAM to the shortcuts of each layer.

Given that CBAM was originally designed to manipulate 2-D data, we modified it to 3-D version according to our dataset. In this module, by using the $l$th layer feature map $\mathbf{F}_l\in \mathbbm{R}^{C\times H\times W\times D}$ as input, CBAM first derives the channel attention module $\mathbf{A}_c\in \mathbbm{R}^{C\times 1\times 1\times 1}$ along the channel dimension and multiplies the feature map with the calculated channel attention map, which can be described as follows:
\begin{equation}
    \mathbf{F'}_l = \mathbf{F}_l\times \mathbf{A}_c(\mathbf{F}_l)
\end{equation}

Then the spatial attention module $\mathbf{A}_s\in \mathbbm{R}^{1\times H\times W\times D}$ along the spatial dimension is connected, also we multiply the computed spatial attention map with the input feature maps, as follows: 
\begin{equation}
    \mathbf{F''}_l=\mathbf{A}_s(\mathbf{F'}_l)  \times\mathbf{F'}_l
\end{equation}

By adding the CBAM module to the shortcuts, the network can force the learning process opt to discard the possible redundant information and hence lead to a better reconstruction result.

\subsection{Loss function}
In the design of the loss function, we take into account the loss from both image and frequency domains. In the image domains, to reduce the blur regions during reconstruction, we choose the mean absolute error (MAE) as our reconstruction loss, as follows:
\begin{equation}
    \mathcal{L}_{recon} = \mathbbm{E}_{x,y,z}|| G-V ||_1
\end{equation}
where $x,y$ and $z$ are the voxel indexes along x, y and z axes respectively.

As proposed in section 3.2, the EAM provides rich structural and textural details of the edge region, which can contribute more to the 3-D accurate reconstruction. Hence, considering that edge regions only occupy a small proportion in the computing of the overall reconstruction loss, we also adopt the output of the edge-aware attention module to design a loss focused on the edge regions. To be more specific, we calculate the binary cross-entropy (BCE) between the extracted edges from ground truth (GT) and reconstructed vertebrae, respectively. The details are as follows: 

\begin{equation}
\begin{aligned}
    \mathcal{L}_{edge} = -\mathbbm{E}_{x,y,z}[&E(G)\ln(E(V)+\\&(1-E(G))\ln(1-E(V)]
\end{aligned}
\end{equation}
where $G$ represents GT vertebrae, $V$ represents reconstructed vertebrae and $E(\cdot)$ denotes extract edge operation.

To further improve the reconstruction accuracy of the edge regions, we take the statistical properties of high-frequency information from the frequency domains into account. In this way, the weight of the loss caused by the edge regions can be increased which can make the network pay more attention to the edge, contour and the corresponding texture information. To evaluate the loss, we define the loss term using the Euclidean distance, as follows:
\begin{equation}
\begin{aligned}
    &\mathcal D_{real} = \mathbbm{E}_{x,y,z}||(\mathcal R(G)-\mathcal R(V))||_2  \\
    &\mathcal D_{imag} = \mathbbm{E}_{x,y,z}||(\mathcal I(G)-\mathcal I(V))||_2
\end{aligned}
\end{equation}
where $\mathcal R$ and $\mathcal I$ denote the real and image components of the feature maps in the frequency domain after performing the FFT, respectively.

Furthermore, the mean value calculation potentially causes the weakening of high-frequency information. Hence, we still retain the original Euclidean distance but utilize the logarithm calculation to reduce the magnitude of the excessive accumulation in this loss term. Here, we name the loss term as the frequency distance (FD), as follows:

\begin{equation}
    \mathcal{L}_{freq} = \log_{10}(\mathcal D_{real}+D_{imag})
    \label{eqa7}
\end{equation}

Considering that MIP cannot only preserve the whole shape of structures but also the relative position of different structures, we compute the MIP both of the reconstructed and the real vertebrae from three orthogonal views, respectively. In this procedure, the three views are consistent with the conventional direction in the medical imaging community (axial, coronal, and sagittal). In this study, we utilize the L1 norm to evaluate the similarity of the MIP images. The projection loss is defined as follows:
\begin{equation}
\begin{aligned}
    \mathcal{L}_{proj} = \frac{1}{3}\big[& \mathbbm{E}_{x}|| \max (G_x(y,z))-\max ( V_x(y,z))||_1 +\\& \mathbbm{E}_{y}|| \max (G_y(x,z))- \max (V_y(x,z))||_1
    +\\&\mathbbm{E}_{z}||\max (G_z(x,y))- \max (V_z(x,y))||_1\big]
\end{aligned}
\end{equation}

Finally, the total objection loss function can be summarized as follows:
\begin{equation}
    \mathcal{L}_{total} = \lambda_1\mathcal{L}_{recon}+\lambda_2\mathcal{L}_{edge}+\lambda_3\mathcal{L}_{freq}+\lambda_4\mathcal{L}_{proj}
\end{equation}
where $\lambda_1$, $\lambda_2$, $\lambda_3$, and $\lambda_4$ are the weights controlling the contribution of different loss terms.

\section{Experiments}
\subsection{Datasets}
To evaluate the proposed method, it is necessary to provide a dataset with bi-planar X-rays and their corresponding CT volumes. Obtaining such a dataset is challenging and constructing one from scratch is impractical. As an alternative, we adopt the following strategy: we first select three publicly available datasets to synthesize the corresponding simulated X-ray images using DeepDRR \cite{MathiasUnberath2018}. This technique has been proven to be effective in prior works \cite{gao2022,killeen2023}. More specifically, we validate the broad applicability of our proposed EAR for bi-planar 3-D reconstruction on those datasets: Verse19 \cite{loffler2020}, VerSe20 \cite{liebl2021,sekuboyina2021,glocker2013}, CTSpine1K \cite{Deng2021}. These datasets are used for vertebrae segment task. These datasets contain the original CT scans and masks for each vertebrae. We first extract individual vertebrae from CT scans using the offered masks and resample them to a $1\times 1\times 1$ mm$^3$ resolution. These resampled vertebrae scans serve as the GT. The 2-D inputs are then constructed by synthesizing DRRs from these GT volumes.

The accurate reconstruction of vertebrae highly depends on the correspondence of features from multiple views. However, during the imaging process of vertebrae, variations in patient pose or the position of the hospital bed can decrease the reconstruction quality due to partial overlapping of the same structure. To mitigate potential interference from these imaging challenges, we utilize the DeepDRR model to obtain projected images from different imaging angles, focusing solely on the influence of the network and loss terms on reconstruction accuracy. To improve the generalization ability of the proposed method, we utilize three public datasets, including Verse19 \cite{loffler2020}, VerSe20 \cite{liebl2021,sekuboyina2021,glocker2013} and CTSpine1K \cite{Deng2021}. The details are as follows:

\textbf{VerSe19} is a vertebral labeling and segmentation dataset prepared for a grand challenge hosted at the 2019 Medical Image Computing and Computer Assisted Intervention (MICCAI), including 160 CT scans.

\textbf{VerSe20} is also prepared for a grand challenge hosted at the 2020 MICCAI, containing 319 scans from 300 patients.

\textbf{CTSpine1K} is collected by the Key lab of Information Processing of the Chinese Academy of Sciences, which contains 807 public CT scans. 

After removing the duplicated or incomplete scans, we eventually extracted 5838 vertebrae as the dataset. The dataset is divided into training set, validation set and test set according to a ratio of 8:1:1.

\subsection{Evaluation Metrics}
To quantitatively evaluate the proposed method and conduct the comparison experiments, we adopt six different metrics, including mean square error (MSE), MAE, Dice coefficient (Dice), peak signal noise ratio (PSNR), structural similarity index measure (SSIM) and FD. The generation quality and difference between the reconstruction and GT can be measured by calculating these metrics.

\textbf{MSE} and \textbf{MAE} are both utilized to evaluate the proximity of the reconstructed results and GT, but utilize L1 and L2 norm, respectively. In this way, we can pay more attention to the outlier removal. The closer to 1 means the higher reconstruction accuracy. In detail, MSE is computed as follows:
\begin{equation}
    MSE = \mathbbm{E}_{x,y,z} ||GT - pred||_2
\end{equation}

\textbf{Dice} can be utilized to measure the overlap ratio between the predicted results and GT. The range of Dice coefficient is $[0,1]$. The closer to 1 means the higher reconstruction accuracy while closer to 0 means the lower accuracy. The Dice can be computed as follows:
\begin{equation}
    Dice = \frac{2\times (pred \bigcup GT)}{pred \bigcap GT}
\end{equation}
where $pred$ denotes the predicted results, while $\bigcup$ denotes the intersection set and $\bigcap$ denotes the union set.


\textbf{PSNR} can be utilized to evaluate the quality of the reconstructed results by calculating the ratio of the energy of the peak signal to the average energy of the noise. A higher PSNR indicates a higher percentage of signal components in the reconstruction results with better quality.
\begin{equation}
    PSNR = 10\times \log_{10}({\frac{Max Values^2}{MSE}})
\end{equation}
Since the CT is often recorded with 12 bits, we set the $Max Values$ to 4096.

\textbf{SSIM} can be exploited to objectively evaluate the quality from the view of the human visual system. The value of SSIM is between $[-1,1]$, and the closer to 1 indicates that the result is more consistent with the high quality of human vision as well as high similarity. SSIM can be computed as follows:
\begin{equation}
    SSIM = \frac{(2\mu_{pred}\mu_{GT}+C_1)(2\sigma_{{pred}{GT}}+C_2)}{(\mu_{pred}^2+\mu_{GT}^2+C_1)(\sigma_{pred}^2+\sigma_{GT}^2+C_2)}
\end{equation}
where $\mu$ denotes mean value, $\sigma$ denotes variance, $C1$ and $C2$ is added to prevent dividing by zero.

\textbf{FD} is designed as a metric to evaluate the accuracy of the reconstructed results concerning the edge regions since edge information is crucial for preserving the structure of spines. Theoretically, the range of FD is in $[0,+\infty]$, where a lower value indicates closer proximity to the GT in the frequency domain. To enhance result comparability, we apply a logarithmic operation and min-max normalization to the FD, resulting in a range of $[0,1]$. A lower value now signifies a higher similarity in the frequency domain.

\begin{figure*}[htbp]
    \centering
    \includegraphics[width=\textwidth]{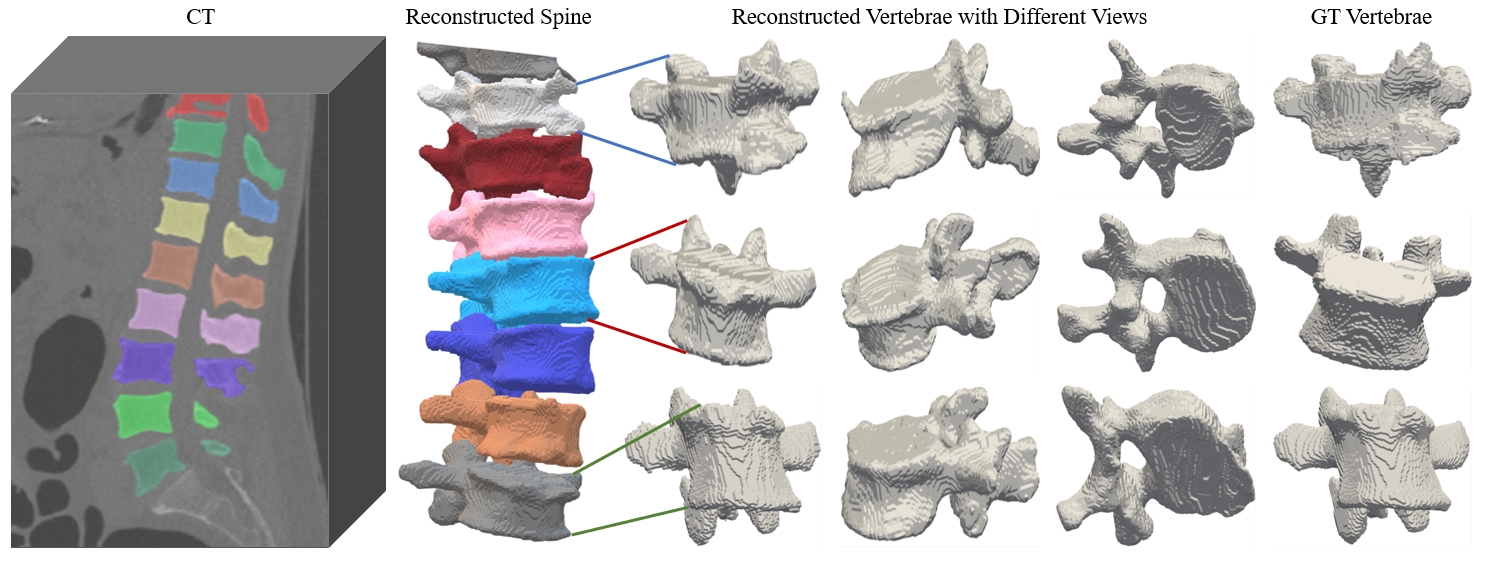}
    \caption{Example of the reconstructed spine and single vertebrae from different locations of the spine.}
    \label{fig_spine}
\end{figure*}

\subsection{Methods for comparison}
To evaluate the performance of the proposed method, we compare it with four state-of-the-art (SOTA) methods: PSR \cite{LiyueShen2019}, Single Image Tomography (SIT) \cite{PhilippHenzler2018}, End-to-End reconstruction (E2E) \cite{YoniKasten2020} and BX2S-Net (BX2S) \cite{ZheyeChen2023}. We achieve PSR and BX2S using public implementation. We re-implemented SIT and E2E in strict accordance with the original papers. The vanilla SIT is designed for single view input, we modified it to support bi-planar input. Specifically, we changed the first layer of SIT to a convolution layer with an input channel of 2. The E2E is originally designed for bi-planar input. Therefore, we reproduce the E2E as a 3-D auto-encoder with a dimensionality enhancement following \cite{YoniKasten2020}. To be fairly compared, all the models are trained using the same parameters and settings.

\subsection{Implement details}
We implemented all the algorithms using PyTorch1.12 under the Ubuntu 20.04 environment and conducted all the experiments on a local computer with an NVIDIA GeForce RTX 3090 GPU and 2.4GHz Intel CPU. We use Adam as the optimizer, with an initial learning rate of 0.01, a batch size of 2, and a weight decay rate of 0.00001. A cosine annealing learning rate policy with a linear warmup for the first 10 epochs is used. The maximum epoch is 100. In addition, the $\lambda_1$,$\lambda_2$,$\lambda_3$ and $\lambda_4$ in objective function is empirically set as 1, 0.1, 0.1 and 0.1. For a fair experimental comparison, all the comparative networks are trained using the same setting.

\section{Experimental results}
To demonstrate the effectiveness of the proposed model, we first visualized the reconstructed results of a spine, as shown in \cref{fig_spine}. The prediction results are CT volumes. To better visualize the reconstruction results, we used the Marching Cubes algorithm to extract meshes from the predicted volumetric results. In this figure, the image on the first column refers to the original CT volume of the spine. Structures in the second column are the vertebrae from top T10 to bottom L5 which can be distinguished by the masks with different colors. The reconstructed vertebrae are shown from the third to fifth columns in different view angles. It can be seen that the proposed method is capable of reconstructing the stitched spine. In each row of the figure, we display the reconstructed results of  T11 thoracic vertebrae, L2 and L5 lumbar vertebrae, respectively. It can be observed that the proposed method accurately reconstructs the shapes and local details of different vertebrae. In particular, as illustrated by T11 vertebrae, our method also demonstrates its possibility even when dealing with the local vertebral deformities.

\begin{figure*}[htbp]
    \centering
    \includegraphics[width=0.9\textwidth]{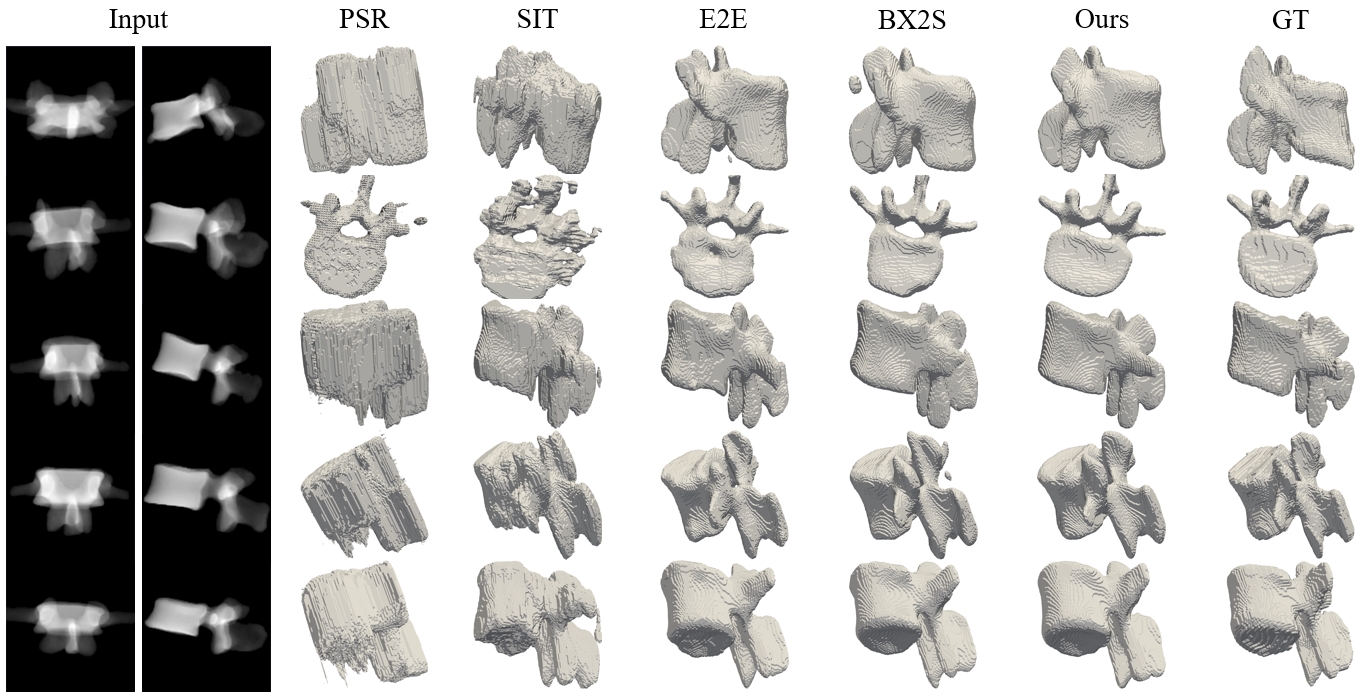}
    \caption{Visualization of vertebrae models reconstructed from different methods: Patient-Specific Reconstruction (PSR), Single Image Tomography (SIT), End-to-End (E2E), BX2S-net (BX2S) and Ours. GT refers to ground truth.}
    \label{fig8}
\end{figure*}
 
\begin{figure}[htbp]
    \centering
    \includegraphics[width=\linewidth]{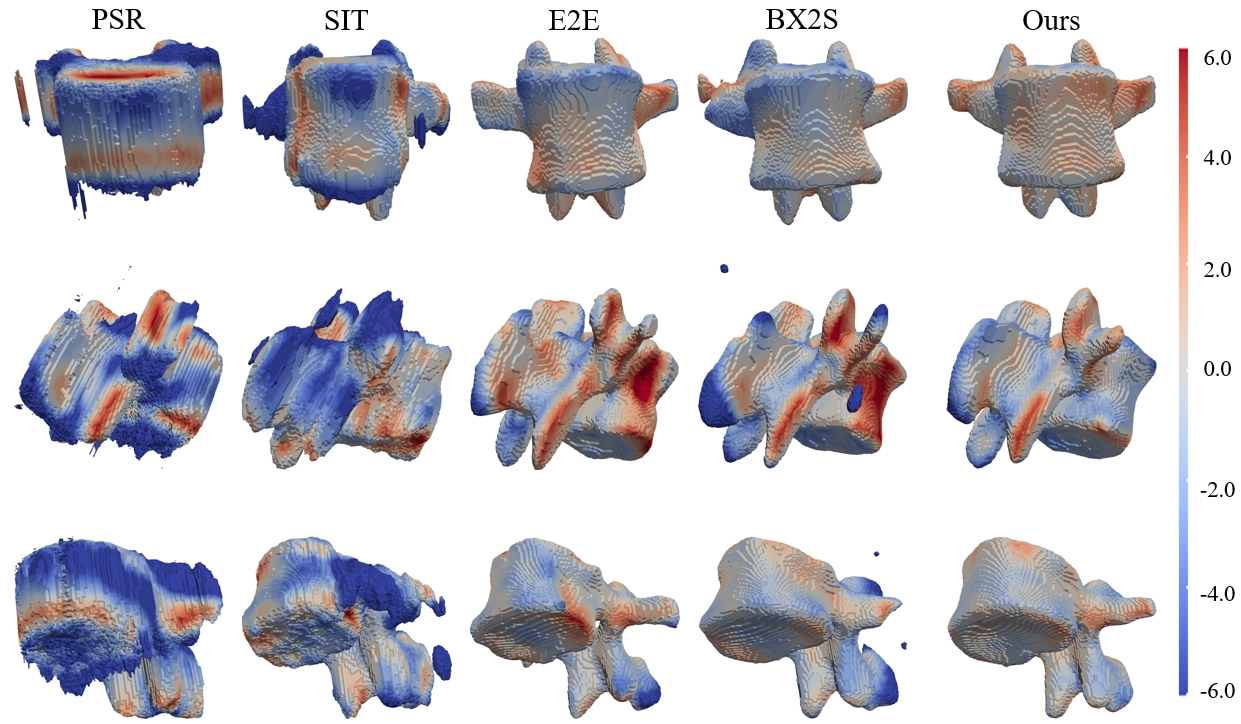}
    \caption{Surface distance error between the output from different methods: Patient-Specific Reconstruction (PSR), Single Image Tomography (SIT), End-to-End (E2E), BX2S-net (BX2S) and Ours.}
    \label{fig9}
\end{figure}

To better visualize the superiority of EAR, \cref{fig9} shows the heat maps by computing the surface distance error (SDE) between GT and reconstructed results by different methods, respectively. In the visual results, the PSR reconstructs the vertebral body and spinous process as a cylinder, resulting in large deviations. The SIT fails to handle uneven areas. The E2E and BX2S exhibit poor performance in morphological irregular areas such as the protrusions at the bottom of the spinous process and the depressions on the lateral sides of the vertebral body. In addition, due to the introduction of FFAG, the BX2S tends to generate discrete noise. In contrast, the proposed EAR achieved better reconstruction results in recovering the sharp regions and maintaining detailed vertebrae structures.

To quantitatively compare the proposed method with other methods, \cref{table1} presents the results by computing MSE, MAE, Dice, PSNR, SSIM and FD. The proposed EAR outperforms the other methods concerning all six metrics. Specifically, the proposed method achieves 25.32\%, 15.32\%, 86.44\%, 23.7612, 80.13\% and 0.3014 in terms of MSE, MAE, Dice, PSNR, SSIM and FD, respectively, which are 4.74\%, 2.09\%, 2.44\%, 0.9633, 1.72\% and 0.0587 higher than other comparative methods. The PSR shows better performance in terms of the Dice score while it performs the worst on other metrics. The possible reason is the cylinder assumption makes the overlap area between the GT and reconstructed results larger. In particular, the Dice score of BX2S is 3.74\% lower than that of EAR, which further demonstrates the contribution of the EAM and FEM to edge preservation.

\begin{table*}[htbp]
    \centering
    \caption{Quantitative comparison results of Ours and other methods in terms of MSE, MAE, Dice, PSNR, SSIM and frequency distance (FD).}
    \resizebox{\linewidth}{14.4mm}{
        \begin{tabular}{c c c c c c c}
        \hline
            \textbf{Method} & \textbf{MSE} & \textbf{MAE} & \textbf{Dice} & \textbf{PSNR} & \textbf{SSIM} & \textbf{FD}    \\
            \hline
            PSR &0.4393$\pm$ 0.0027 & 0.4642$\pm$ 0.0007 &0.8161$\pm$ 0.0112& 21.1509$\pm$ 0.8415& 0.2041$\pm$ 0.0012&0.4879$\pm$ 0.0006  \\
            SIT &0.4118$\pm$ 0.0004&0.2863$\pm$ 0.0001&0.4451$\pm$ 0.0004&21.2628$\pm$ 0.2196&0.6464$\pm$ 0.0002&0.9525$\pm$ 0.0370 \\
            E2E &0.3032$\pm$ 0.0101&0.1764$\pm$ 0.0016&0.8400$\pm$ 0.0016&22.9527$\pm$ 2.7255&0.7831$\pm$ 0.0013&0.3591$\pm$ 0.0047  \\
            BX2S &0.3006$\pm$ 0.0049&0.1743$\pm$ 0.0006&0.8270$\pm$ 0.0010&22.7980$\pm$ 0.7383&0.7823$\pm$ 0.0002&0.7333$\pm$ 0.0119 \\
            Ours &\textbf{0.2532$\pm$ 0.0332}&\textbf{0.1532$\pm$  0.0032}&\textbf{0.8644$\pm$ 0.0354}&\textbf{23.7613$\pm$  1.7245}&\textbf{0.8013$\pm$ 0.0342}&\textbf{0.3014$\pm$ 0.0741} \\
            \hline
        \end{tabular}
    }
    \label{table1}
\end{table*}

\subsection{Ablation Study of the Network Modules}

\begin{figure}[htbp]
    \centering
    \includegraphics[width=\linewidth]{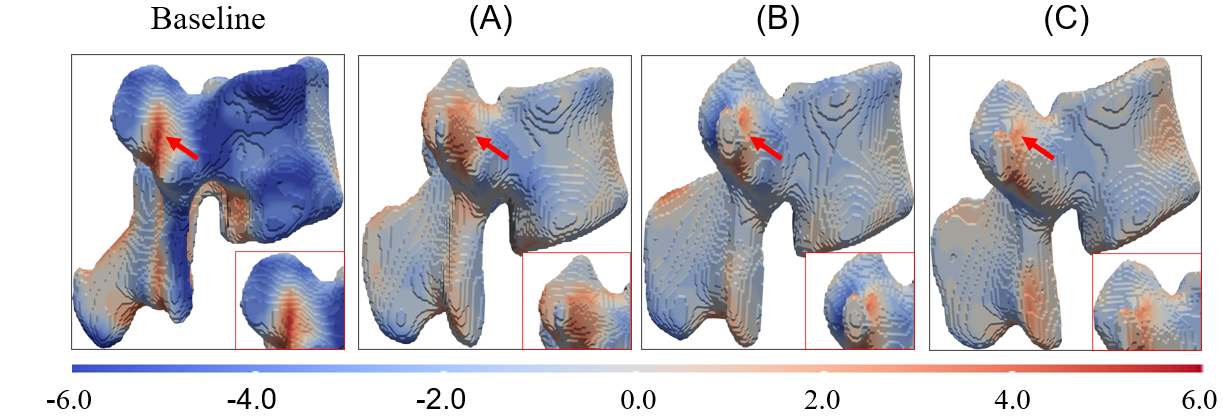}
    \caption{Ablation study of our EAR. The 'Baseline' denotes the auto-encoder backbone. (A) adding attention-based shortcut to the auto-encoder, (B) incrementally adding frequency enhancement module, (C) incrementally adding edge attention module based on (B). The setup of experiment (C) denotes the full of EAR.}
    \label{fig_ablation}
\end{figure}

\begin{figure}[htbp]
    \centering
    \includegraphics[width=\linewidth]{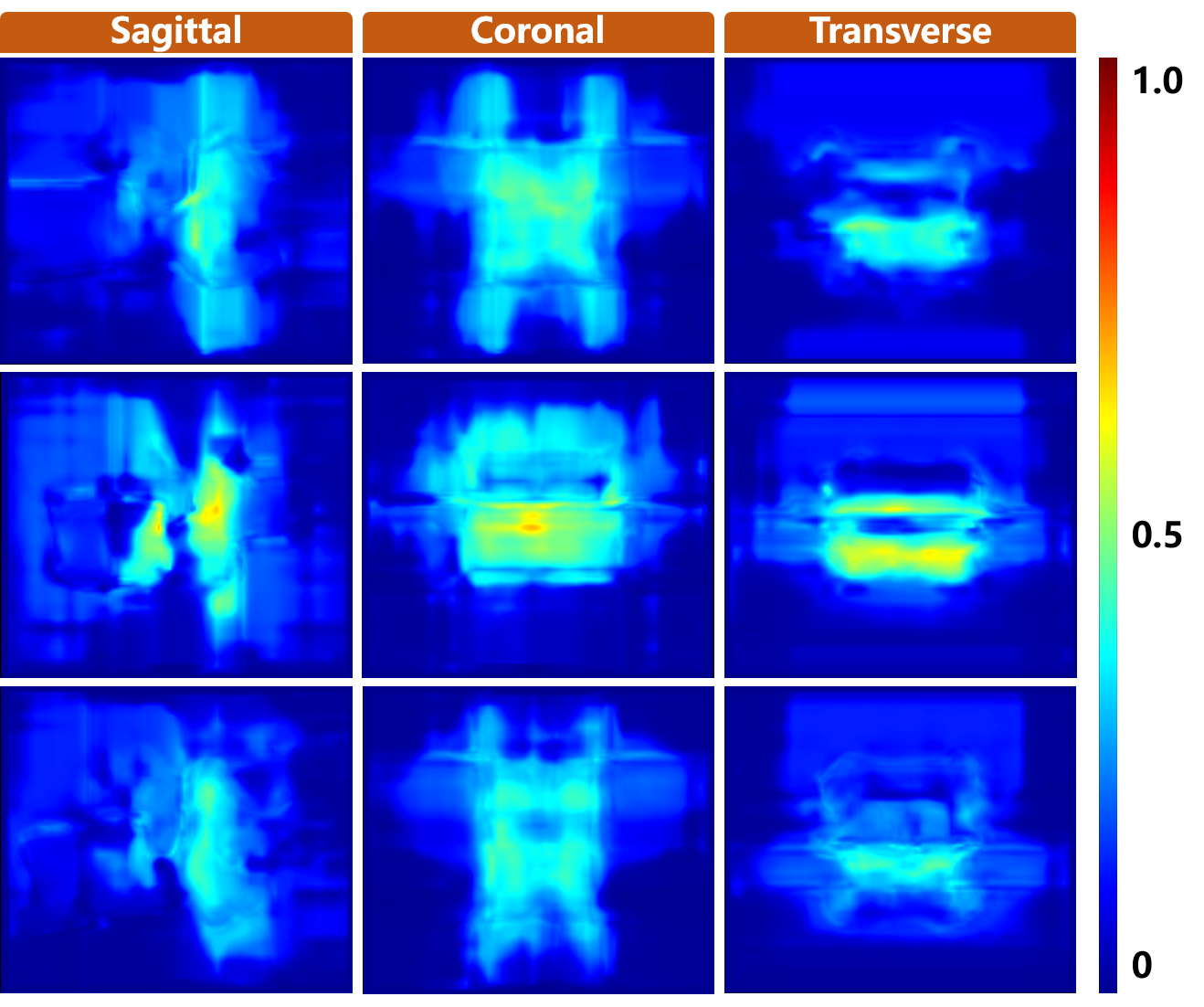}
    \caption{Visualization of the Edge Attention Module for three randomly selected vertebrae.}
    \label{eam_vis}
\end{figure}

\begin{figure}[htbp]
    \centering
    \includegraphics[width=0.8\linewidth]{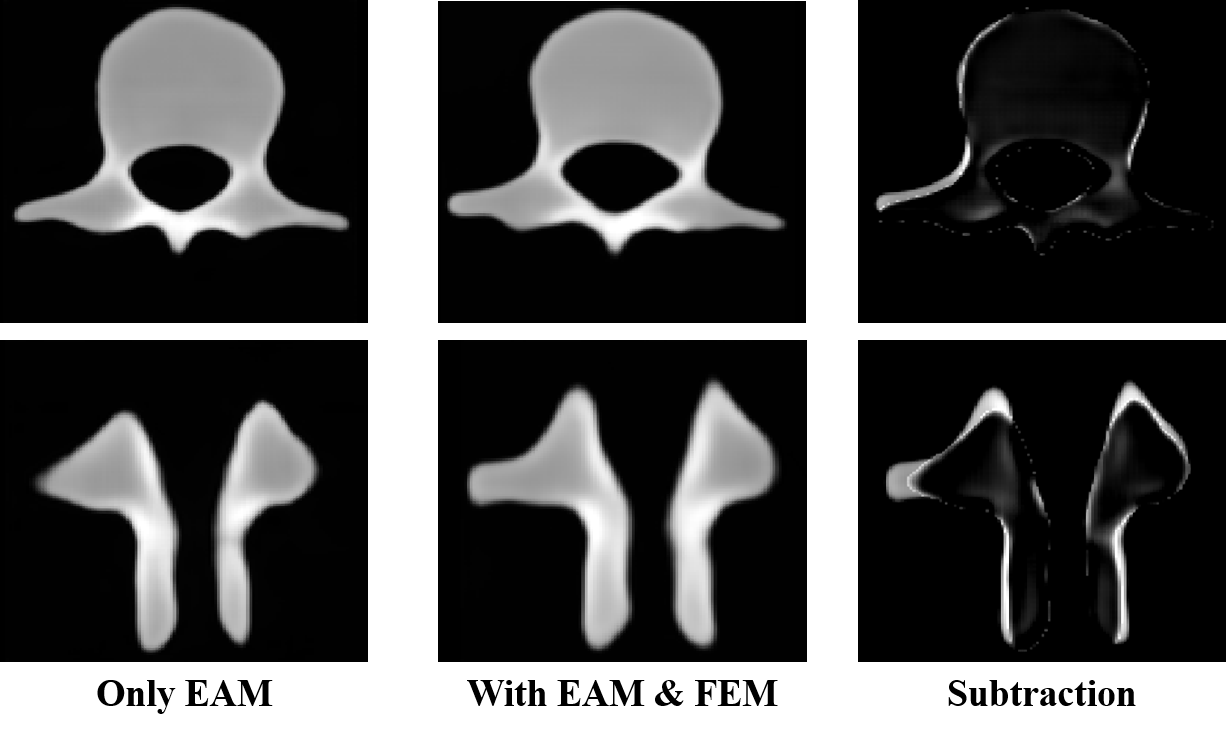}
    \caption{Subtraction result by subtracting prediction with Frequency Enhancement Module (FEM) from prediction without FEM.}
    \label{freq_vis}
\end{figure}

To verify the effectiveness of the proposed modules, we conducted ablation studies based on Baseline. The results are illustrated in \cref{table2} and \cref{fig_ablation}, while the visualization results of the proposed modules are demonstrated in \cref{eam_vis} and \cref{freq_vis}, from which we draw several conclusions.

The comparison results in \cref{table2} show that ABS improves the reconstruction results by 0.73\%, 8.46\% in terms of MSE and Dice, which shows that ABS benefits from removing redundant noises and improving reconstruction. We also visualize the reconstructed results in \cref{fig_ablation}. After adding ABS, the reconstruction results are displayed in the second column. Lighter colors on the reconstructed surface indicate a closer match to the GT.

Based on the above network, we also present the results in \cref{table2} after adding the FEM. 
It is observed that FEM improves the results with MSE, MAE, PSNR and SSIM increasing by 1.37\%, 0.18\%, 1.0597, 0.18\%, respectively, which preliminarily explains that the FEM obtains more detailed sharp structures. Moreover, FD is significantly improved. We also further visualize reconstructions of the network with FEM in \cref{fig_ablation}. Thanks to FEM, the color of the sharp structure is relatively light, which implies that FEM tends to discover the high-frequency component of the vertebrae and adequate edge information is captured. 

In the evaluation of EAM, the comparison results in \cref{table2} show that EAM improves the reconstruction performance by 0.27\%, 1.47\% and 0.3217 in terms of MSE, Dice and PSNR, respectively, demonstrating better edge reconstruction ability of the network with EAM. As shown in \cref{fig_ablation}(C), both the red and blue areas are lighter than the previous results, indicating that the EAM contributes to a more accurate reconstruction. 

We further visualize feature maps in the proposed EAM, as show in \cref{eam_vis}. Three samples were randomly selected to visualize the edge attention maps from the sagittal, coronal and transverse views. Brighter regions indicate a high probability of being identified as an edge region. With EAM, the surface part of the vertebrae appears particularly bright, indicating that EAM tends to detect the edge part and capture richer spatial information.

To further verify that FEM can improve the accuracy of high-frequency edge region and clarify the necessity of combining the FEM and EAM, we visualize the effect of FEM. Specifically, we randomly select a sample and predict the results using models with only EAM and with both EAM and FEM, respectively. Then we visualize the difference by subtracting the results without FEM from the results with FEM. The subtraction result is shown in \cref{freq_vis}. It can be observed that after subtraction, residual parts mainly concentrate in the high-frequency edge regions, indicating that incorporating FEM results in finer and more complete edge reconstruction, validating the effectiveness of FEM.

\begin{table*}[htbp]
    \centering
    \caption{Evaluation of the proposed components. 'Baseline' denotes the vanilla auto-encoder, 'ABS' denotes the attention-based shortcuts, 'FEM' refers to the frequency enhancement module and 'EAM' means the edge attention module.}
    \label{table2}
    \resizebox{\linewidth}{12mm}{
        \begin{tabular}{c c c c c c c c c c}
            \hline
            \textbf{Baseline} & \textbf{ABS} & \textbf{FEM} & \textbf{EAM} & \textbf{MSE} & \textbf{MAE} & \textbf{Dice} & \textbf{PSNR} & \textbf{SSIM} & \textbf{FD}    \\
            \hline
            \checkmark & & & & 0.2769$\pm$ 0.0087 & 0.1551$\pm$ 0.0010 & 0.8502$\pm$ 0.0028 & 23.1218$\pm$ 0.6866 & 0.8035$\pm$ 0.0005 & 0.7103$\pm$ 0.0012 \\
            \checkmark & \checkmark & & & 0.2696$\pm$ 0.0147 & 0.1508$\pm$ 0.0147 & 0.8666$\pm$ 0.0253 & 22.3799$\pm$ 0.8139 & 0.8046$\pm$ 0.0101 & 0.7019$\pm$ 0.1173 \\
            \checkmark & \checkmark & \checkmark &  &0.2559$\pm$ 0.0024&\textbf{0.1490$\pm$ 0.0002}&0.8514$\pm$ 0.0006&23.4396$\pm$ 0.6898&\textbf{0.8068$\pm$ 0.0002}&0.3125$\pm$ 0.0960  \\
            \checkmark & \checkmark & \checkmark & \checkmark  &\textbf{0.2532$\pm$ 0.0076}&0.1532$\pm$  0.0011&\textbf{0.8661$\pm$ 0.0012}&\textbf{23.7613$\pm$ 2.8550}&0.8013$\pm$ 0.0011&\textbf{0.3014$\pm$ 0.0053} \\
            \hline
        \end{tabular}
    }
\end{table*}

\subsection{Ablation Study of Loss Function}

The performances of different loss terms are summarized in \cref{table_loss}. We set the reconstruction loss as the base loss following other methods \cite{PhilippHenzler2018, LiyueShen2019, YoniKasten2020} and gradually add loss terms to evaluate the effectiveness of each loss term. As the first row shows, reconstruction loss performs the worst on MSE and FD due to the lack of edge constraints. The edge loss contributes slightly to improving the results. After adding the frequency loss, significant improvement has been made by constraining the high-frequency feature similarity and hence can lead the network to reconstruct the sharp edge features. The projection loss brings additional improvement in the reconstruction quality by constraining the shape contours of the reconstructed volume from three axes during training. Therefore, the projection loss can improve the shape consistency. From the table, the values of PSNR, SSIM and FD of the network with all the loss terms outperform all the other combinations. The results indicate that the projection loss improves the shape similarity of the prediction and GT. 

\begin{table*}[htbp]
    \centering
    \caption{Evaluation of different loss settings during training. '$\mathcal L_{recon}$', '$\mathcal L_{edge}$' and '$\mathcal L_{freq}$' denote the reconstruction loss, edge loss, and frequency loss respectively. '$\mathcal L_{mip}$' means that the projection loss is enabled.}
    \resizebox{\linewidth}{12mm}{
        \begin{tabular}{c c c c c c c c c c}
        \hline
            \textbf{$\mathcal L_{rec}$} & \textbf{$\mathcal L_{edge}$} & \textbf{$\mathcal L_{freq}$} & \textbf{$\mathcal L_{proj}$} & \textbf{MSE} & \textbf{MAE} & \textbf{Dice} & \textbf{PSNR} & \textbf{SSIM} & \textbf{FD}    \\
            \hline
            \checkmark & & & &0.2910$\pm$ 0.0340&0.1572$\pm$ 0.0340&0.8615$\pm$ 0.0337&23.2530$\pm$ 1.5946&0.7915$\pm$ 0.0316&0.7663$\pm$ 0.0639 \\
            \checkmark & \checkmark & & &0.2892$\pm$ 0.0122&0.1541$\pm$ 0.0122&0.8638$\pm$ 0.0225&22.9221$\pm$ 0.7784&0.7956$\pm$ 0.0099&0.7244$\pm$ 0.1149 \\
            \checkmark & \checkmark & \checkmark & &\textbf{0.2488$\pm$ 0.0160}&\textbf{0.1470$\pm$ 0.0160}&\textbf{0.8682$\pm$ 0.0215}&23.5782$\pm$ 0.9134&0.7914$\pm$ 0.0146&0.3367$\pm$ 0.0680  \\
            \checkmark & \checkmark & \checkmark & \checkmark &0.2532$\pm$ 0.0332&0.1532$\pm$  0.0032&0.8661$\pm$ 0.0332&\textbf{23.7613$\pm$  1.7245}&\textbf{0.8013$\pm$ 0.0342}&\textbf{0.3014$\pm$ 0.0741}  \\
            \hline
        \end{tabular}
    }
    
    \label{table_loss}
\end{table*}


\section{Discussion}
In this study, we proposed a novel and robust edge-aware reconstruction method. From a methodological perspective, while previous 3-D reconstruction methods focus only on the distribution of voxels within CT volumes, our approach prioritizes edge information. The proposed reconstruction method was evaluated by using three public datasets comprising 5838 vertebrae from 1162 CT scans. A comparative quantitative validation was conducted to assess whether our proposed technique is superior to previously developed PSR, SIT, E2E, and BX2S. In PSR, the 2-D inputs are directly fed into the network. This may slow down convergence, cause false reconstructions, and introduce significant noise. SIT, on the other hand, smooths out many high-frequency features during the 2-D to 3-D transformation, resulting in inaccuracies in reconstructing uneven areas of the vertebrae. As for E2E, while dimension expansion reduces the dimension gap, it overlooks edge information from the vertebrae, thereby inadequately handling complex and asymmetrical vertebrae surfaces. Regarding BX2S, the FFAG may guide the network's attention to redundant parts brought by dimension enhancement, leading to noise outside the vertebrae. In the validation of the reconstruction performance of the proposed EAR, we obtained MAE of 0.1532 and PSNR of 23.7613 for test sets. The results indicate that incorporating FEM and EAM can improve the reconstruction accuracy of sharp edge features. Previous works have attempted to improve edge constraint by adding an edge channel of the output to predict the edge of the reconstructions \cite{ZheyeChen2023}. In contrast, we utilized an edge loss term to directly enforce edge similarity. Moreover, the designed frequency loss term is as substantial to the sharp edge region as edge loss. 




\begin{figure}[htbp]
    \centering
    \includegraphics[width=\linewidth]{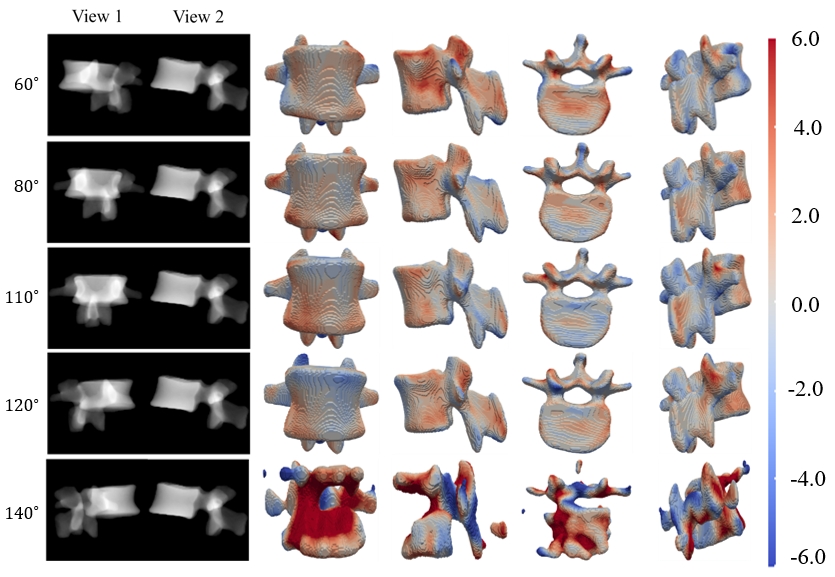}
    \caption{Visualization of reconstruction results from non-orthogonal bi-planar digitally reconstructed radiographs (DRRs). View 1 and view 2 denote the first and second views of the bi-planar DRRs we use. View 1 is changeable and view 2 is fixed at the lateral view. The number on the left tells the angle between view 1 and view 2.}
    \label{fig_limit}
\end{figure}

To study the potentiality of the proposed method when the angle of the bi-planar X-rays varies, especially in the situation of non-strictly orthogonal, we trained the proposed network when the angle of two views gradually increases from 60$^\circ$ to 140$^\circ$, as shown in \cref{fig_limit}. As can be seen from the figure, when the angle is $60^\circ$, the reconstruction exhibits a lot of red in the SDE, indicating that the reconstructed surface is mostly inside the GT surface. When the angle is $80^\circ$ and $110^\circ$, the overall color of the reconstruction is lighter, which denotes an acceptable reconstruction. As the angle increases to $120^\circ$, the color of the reconstructed vertebrae processes is darker, indicating a decrease in the reconstruction performance of the edge region. When the angle reaches $140^\circ$, since the X-rays are fluoroscopic, the non-overlapping information provided by the two views is greatly reduced, resulting in wrong reconstructions.

\section{Conclusion}
This study tackles the challenging task of reconstructing 3-D structures from bi-planar X-ray images using deep neural networks, particularly when faced with limited perspective information. The proposed EAR can achieve sparse-view 3-D reconstruction, directly recovering lost 3-D spatial information from 2-D X-ray images taken at limited viewing angles, and compute relatively clear vertebral morphologies. The proposed EAR is automated and has a very short inference time, making it highly suitable for real-time applications, including intraoperative reconstruction. Given that EAR prioritizes edge information and enhances the recovery of edge details from both spatial and frequency domains, it achieves a more precise reconstruction of asymmetric and uneven areas of the vertebral surface, thereby providing support for clinicians to better identify vertebral lesions. This achieves the challenge of inaccurate prediction in edge regions existed in current methods. Specifically, to enhance feature fusion between encoder and decoder layers, we incorporate the CBAM into the shortcuts. Additionally, we also propose the EAM to extract crucial edge information from shallow layers, aiding in the reconstruction process’s accuracy, especially concerning edge details. The versatility of our proposed EAM allows it to serve as a plug-in module for edge-related learning across various 3-D reconstruction tasks. Furthermore, we devise the FEM to amplify network attention towards high-frequency information. Empirical evidence demonstrates that EAR outperforms SOTA methods such as PSR, SIT, E2E and BX2S in terms of MSE, MAE, Dice, PSNR and SSIM. These results underscore the significant potential of our method for accurate 3-D reconstruction from bi-planar X-ray images.

However, the limitations and challenges are as follows: First, with regard to X-ray images, the information provided represents a perspective structure within the human body. When the two X-ray image angles input into EAR are not orthogonal, the non-overlapping information that can be provided is significantly reduced. This situation results in the network's inability to accurately infer the 3-D structure of the overlapping parts. Since diffusion model-based methods can generate multi-view consistent images from single-view inputs \cite{Zhou_2023_CVPR}, our future work will explore the extension of diffusion model-based methods to generate dense multi-view X-ray images. Secondly, our method is based on explicit voxel training, which requires significant memory, and the model requires fully supervised training, thus limiting its generalization capability. Therefore, future work will explore extensions based on more effective 3-D representation techniques like Neural Radiance Field (NeRF) \cite{muller2022instant} and 3-D Gaussian Splatting \cite{kerbl20233d}, which eliminate the need for training neural network models and instead enable direct reconstruction of the target 3-D volume through iterative optimization. \cite{zhao2018parallel}.

\printcredits

\appendix
\section{Declaration of Competing Interest}
The authors declare that they have no known competing financial interests or personal relationships that could have appeared to influence the work reported in this paper.

\section{Acknowledgement}
This work was supported in part by grants from National Key Research and Development Program of China
(2023YFB4706003), National Natural Science Foundation of China (62176268), and Beijing Natural Science Foundation-Joint Funds of Haidian Original Innovation Project (L232022).

\bibliographystyle{model1-num-names}

\bibliography{ref}

\end{CJK}
\end{document}